\documentclass[a4paper, amsfonts, amssymb, amsmath, reprint,  showkeys, nofootinbib, twoside, superscriptaddress]{revtex4-1}
\usepackage[english]{babel}
\usepackage[utf8]{inputenc}
\usepackage[colorinlistoftodos, color=green!40, prependcaption]{todonotes}
\usepackage{amsthm}
\usepackage{mathtools}
\usepackage{physics}
\usepackage{xcolor}
\usepackage{graphicx}
\usepackage[left=18mm,right=18mm,top=35mm,columnsep=15pt]{geometry} 
\usepackage{adjustbox}
\usepackage{placeins}
\usepackage[T1]{fontenc}
\usepackage{lipsum}
\usepackage{csquotes}
\usepackage{siunitx}
\usepackage{caption}
\usepackage{subcaption}
\usepackage[pdftex, pdftitle={Article}, pdfauthor={Author}, colorlinks=true]{hyperref}
\usepackage{multirow}
\usepackage{comment}
\renewcommand{\selectlanguage}[1]{}

\begin{document}

\title{\textcolor{black}{Magnetic field} fluctuations induced decoherence of a diamagnetic nanosphere }

\author{Ruiyun Zhang}
    \affiliation{Leiden Institute of Physics, Leiden University, PO Box 9504, 2300 RA Leiden, Netherlands}

\author{Martine Schut}
    \affiliation{Van Swinderen Institute for Particle Physics and Gravity, University of Groningen, 9747AG Groningen, the Netherlands }
    \affiliation{Bernoulli Institute for Mathematics, Computer Science and Artificial Intelligence, University of Groningen, 9747 AG Groningen, the Netherlands \vspace{1mm}}

\author{Anupam Mazumdar }
    \affiliation{Van Swinderen Institute for Particle Physics and Gravity, University of Groningen, 9747AG Groningen, the Netherlands }

\begin{abstract}
This paper provides a simple derivation of the decoherence rate for a diamagnetic nanoparticle in the presence of \textcolor{black}{fluctuations of the magnetic field in a thermal environment}. Diamagnetic levitation is one of the key techniques for trapping, cooling, and creating a macroscopic quantum spatial superposition in many experiments. It is widely applied in many theoretical and experimental endeavours to test fundamental physics in matter-wave interferometers. To estimate the decoherence rate originating from magnetic-field fluctuations, we use the fluctuation-dissipation theorem. 
We show that our resulting decoherence rate expression is analogous to that of a dielectric material \textcolor{black}{interacting with the electric field component of the background field}; however, it is often relatively suppressed due to the material properties.
\end{abstract}

\maketitle


\section{Introduction}\label{sec:intro}

Diamagnetic levitation is a known concept, verified in laboratories~\cite{simon_diamagnetic_2000,Simon:2001} and with wide-ranging applications in quantum theory and experiment~\cite{Nakashima:2020jnr, simon_diamagnetic_2000,hunter2023diamagnetic,hunter2023sonomaglev,Ashkarran_2024,Nakashima:2020jnr,Hofer:2022chf,chen2022diamagnetic,tian2023feedbackcoolinginsulatinghighq,Qian:2013}. For a review, see~\cite{schilling2021physicsdiamagneticlevitation}.
One particular avenue of diamagnetic levitation is its application in creating macroscopic quantum superpositions~\cite{Elahi:2024dbb, van_de_kamp_quantum_2020}, where the idea is to create a spatial superposition in a diamagnetic trap after cooling it to the quantum ground state of translational and rotational degrees of freedom; see~\cite{Hsu:2016,Slezak_2018}.
One can embed a spin defect in a crystal, such as a 
nitrogen-vacancy centre (NVC) in a neutral nanodiamond; for a review; see~\cite{Doherty_2013}. 
Then, this spin can be manipulated with external magnetic pulses to create a macroscopic spatial quantum superposition as in a matter-wave interferometer~\cite{Greve:2021wil,Wan_2016,Nair:2023ovu,Scala_2013,PhysRevLett.125.023602,Folman:2013,PinoEtAl2016,marshman_constructing_2022,PhysRevLett.125.023602,PhysRevLett.123.083601} 
\textcolor{black}{Such systems have a wide range of applications in creating quantum sensors~\cite{wu:2022rdv,Debuisschert_2021}}, to test fundamental physics beyond the Standard Model physics~\cite{Bose:2022czr,Barker:2022mdz,Elahi:2023ozf,Vinckers:2023grv,Chakraborty:2023kel,PinoEtAl2016}, and last but not least, to test the quantum nature of spacetime in a lab \cite{Bose:2017nin,ICTS}, see also~\cite{Marletto:2017kzi}. 
The latter is the most ambitious programme and aims to witness the quantum nature of gravity through entanglement. 
Of course, attaining ground-state cooling is a big challenge for any nanoparticle~\cite{Deli__2020, Piotrowski_2023}, and, as with any quantum experiment, the system's coherence should be well maintained throughout the experiment.
These quantum experiments require a significant demand for controlling the decoherence and random dephasing. \textcolor{black}{
Previously, electromagnetic fluctuations endured by a metallic nanoparticle have been analysed, see Refs.~\cite{rytov_theory_1959, zheng_review_2014,brevik_fluctuational_2022}. However, their application in decoherence has not been understood before Ref.~\cite{sinha_dipoles_2022}. The latter paper considers the decoherence due to the electric component while assuming that the fluctuations have spatial dependence in a long wavelength limit, i.e. when the wavelength of the fluctuations are larger than the size of the nanoparticle.
}

The aim of this paper is simple: we wish to compute the decoherence rate of a diamagnetic nanodiamond due to magnetic field fluctuations in a long wavelength limit. 
\textcolor{black}{We compute the decoherence rate in a steady state condition, assuming that the time dependence of the electric and magnetic fields is long relative to the experimental time scale. This is justifiable in many experimental setups, where a nanodiamond can be levitated diamagnetically by the external magnetic field, and the superposition is created by the gradient in the magnetic field like in the Stern-Gerlach type setup~\cite{Wan_2016,Nair:2023ovu,Scala_2013,PhysRevLett.125.023602,Folman:2013,PinoEtAl2016,marshman_constructing_2022,PhysRevLett.125.023602,PhysRevLett.123.083601}. Such a setup does not rely on the electric field at all, in fact the nanodiamond is considered to be charged neutral.  }

However, the \textcolor{black}{fluctuations} in the magnetic field still interact with a nanodiamond of a finite size and impart stochastic kicks to it. We wish to understand this momentum transfer in a momentum diffusion picture; see~\cite{Fokker:1914, Joos_Zeh_1985,
Berg-Sorenson_1992,Ghirardi,Balykin_1986,Dalibard_1985,Agarwal:1993,
MILONNI_2023, oxenius_kinetic_2012, schlosshauer, rytov_theory_1959}.
Previous studies~\cite{cheng_long-range_1999, chklovskii_relaxation_1992, Joos_Zeh_1985, nagaosa_experimental_1991, BreuerPetruccione2002} have addressed aspects of the magnetic component of fluctuating fields induced by \textcolor{black}{current fluctuations} and magnetization fluctuations in nearby apparatus and their impact on susceptibility. However, to our knowledge, the \textcolor{black}{magnetic field fluctuations in a thermal bath have not been explicitly computed in the context of decoherence of diamagnetic nanoparticles}.
\textcolor{black}{The decoherence effect investigated in this work is driven by fluctuations of the external magnetic field at a finite temperature; the decoherence rate vanishes in the zero-temperature limit of the environment since the environment cannot carry away quantum information of the system.
At non-zero temperature, however, a thermal environment can carry away which-path information, i.e. cause decoherence~\cite{schlosshauer}.}

The mathematical relationship quantifying the connection between fluctuations and stochastic kicks is called the fluctuation-dissipation theorem~\cite{MILONNI_2023, oxenius_kinetic_2012, schlosshauer}. In this paper, we provide a route that has been followed historically by many in the context of atoms in a laser or a cavity~\cite{oxenius_kinetic_2012, Joos_Zeh_1985}, and recently to demonstrate the decoherence rate of a dielectric nanoparticle in the presence of the electromagnetic vacuum in a thermal bath~\cite{MILONNI_2023, sinha_dipoles_2022}. 
The authors of the latter papers use the potent fluctuations and dissipation theorem to establish their results and compute the decoherence rate by relating their scattering constant to the decoherence rate, following the theory of decoherence; see Refs.~\cite{schlosshauer,Hornberger:2003} for example.

Our result is a zeroth-order approximation; in an experiment, there will be time-dependent effects,  and our current computation will not capture any of those effects. \textcolor{black}{The motion of the nanodiamond is in the non-relativistic regime; hence the time-dependence of the magnetic field on the nanodiamond is minuscule. Time-dependence is also present in the superposition size $\Delta x(t)$, which we have not taken into account; our result could therefore be an overestimation~\cite{Schut:2024lgp}.} 
Nevertheless, our computations have practical applications in limiting the diamagnetic traps such that the magnetic-field \textcolor{black}{fluctuations at finite temperature} are not dominant in decohering the quantum spatial superposition. 
However, first, we will briefly recap the momentum diffusion equation (Sec.~\ref{sec:recap}) and then apply it to the diamagnetic nanodiamond in the presence of a static magnetic field (Sec.~\ref{sec:calc}). 
We will compute the decoherence rate in a long-wavelength limit \textcolor{black}{(i.e. $\textbf{k}\cdot\textbf{r} \rightarrow 0$, suggesting that the wavelength of the magnetic field fluctuations $\lambda =2\pi/{\bf k}$ is longer than the size of the nanodiamond.)} by following the diffusion constant, similarly to Ref.~\cite{sinha_dipoles_2022}.
Finally, we compare our results with its electric field dual.

\section{Recap of momentum Diffusion constant}\label{sec:recap}

\textcolor{black}{The physics can be intuitively understood using the Fokker-Planck equation.}~\cite{Fokker:1914,Joos_Zeh_1985}, which is described by 
a partial differential equation for a distribution function $w (p, t)$ of a nanoparticle; see~\cite{Balykin_1986,Dalibard_1985,Agarwal:1993,
Berg-Sorenson_1992}.
\textcolor{black}{In the momentum space, the time evolution of a particle's probability density is given by}
\begin{equation}\label{eq:pde}
\textcolor{black}{\frac{dw}{dt} = -\frac{\partial}{\partial p} (Fw) +\frac{\partial^2}{\partial p^2}(Dw)}\,.
\end{equation}
Here, $F$ is the force and describes the rate of change of the average momentum $\langle p\rangle$ that is imparted to the nanoparticle in the presence of electromagnetic vacuum: $ F=d\langle p\rangle/dt$.
Moreover, $D$ in Eq.~\eqref{eq:pde} satisfies the following equation, see~\cite{Balykin_1986,Dalibard_1985,Agarwal:1993,Berg-Sorenson_1992}:
\begin{equation}
    \textcolor{black}{2D \Delta t= \lim_{t\rightarrow \infty}\left(\langle p^2\rangle -\langle p\rangle ^2\right)}\,,
\end{equation}
\textcolor{black}{where $\Delta t$ is viewed as the time duration starting from \(t=0\),} and $D$ is known as the momentum-diffusion coefficient as a function of momentum. 
In the large time limit, denoting the variance $\langle p^2\rangle -\langle p\rangle ^2 =\langle \Delta p^2\rangle$ (where $\Delta p= p-\langle p\rangle$), we find:
\begin{equation}
 \textcolor{black}{2D \Delta t = \lim_{t\to\infty} \, \langle \Delta p^2(t)\rangle}\,. 
\end{equation}
The momentum diffusion constant $2D\equiv \langle \Delta p^2\rangle/\Delta t$ is closely related to the scattering constant $\Lambda$ in collisional decoherence models; $2 \Lambda = \langle \Delta k^2\rangle/\Delta t$~\cite{sinha_dipoles_2022,Hornberger:2003,schlosshauer}, where the wavenumber $k=p/\hbar$.
The authors in~\cite{sinha_dipoles_2022} established that the momentum diffusion constant can be related to the decoherence rate for a dielectric nanoparticle in the presence of \textcolor{black}{the electric field in a thermal bath.} Their results matched the established decoherence rate computed by the scattering theory of blackbody radiation, see~\cite{schlosshauer,RomeroIsart2011LargeQS}.
The decoherence rate in the long-wavelength limit \textcolor{black}{(i.e. the wavelength of the environment particle is much larger than the size of the spatial superposition, which is given by $\gamma = \Lambda (\Delta x)^2$, with the spatial superposition size given by $\Delta x$.}

In the present discussion, we will extend their analysis to a diamagnetic material. We will employ a similar approach to estimate the decoherence rate due to random momentum transfer from the \textcolor{black}{magnetic field in a thermal bath}. We will consider the finite-temperature effects as well, which will provide us with interesting limits on the depth of the diamagnetic traps we can employ in the experiment.

\section{Momentum diffusion due to fluctuations in magnetic field }\label{sec:calc}

\textcolor{black}{We begin by considering a nanosphere of radius \( a \) possessing a point-like magnetic dipole moment. We will later generalize the result to a finite-volume case by introducing the concept of magnetic susceptibility and using the Clausius-Mossotti relation. In the most straightforward setting, we will assume that there exists a background magnetic field $\bf B$.}
The force acting on a magnetic dipole moment of a nanosphere, $\mathbf{m}$, in the presence of an external magnetic field $\mathbf{B}$ is given by~\footnote{\textcolor{black} {In general the force will also contain the derivative of the Electric field, $({1}/{c^2})\epsilon_{ijk}m_j \partial_t E_k$, where $\epsilon_{ijk}$ is the Levi-Civita symbol and $\mathbf{E}$ is the external electric field (with components $E_k$, $k=x,y,z$. However, in the experimental setup we are envisaging there is no charge, nor is there any electric field or it's gradient. The experimental setup will only involve the magnetic field and its spatial gradient, see~\cite{Elahi:2024dbb}.}}:
\begin{align}
    F_i = m_j\partial_i B_j
\end{align}
The impulse on the nanosphere due to the external magnetic field is given by:
\begin{align}
    \Delta p_i &= \int_0^{\Delta t} dt F_i
    = m_j\int_0^{\Delta t} dt\, (\partial_i B_j) \, .
\end{align}
We first consider a classical background field, where we take  $\textbf{r}=0$ at the position of the magnetic dipole moment of the nanosphere. 
We also assume the idealised case of a uniformly magnetised sphere with a small radius, $a$, where 
\begin{equation}\label{eq:mj}
    m_j = \alpha B_j
\end{equation}
where $\alpha$ depends on the vacuum permeability $\mu_0$ and the volume and permittivity of the nanocrystal.
We are assuming the background magnetic field in the long-wavelength approximation, which is given by:
\begin{align}
    B_j &=  i\sum_{\mathbf{k},\lambda} \left(\frac{\hbar}{2\epsilon_0\omega V}\right)^{1/2} \left[ \hat{a}_{\mathbf{k},\lambda}e^{-i\omega t}- \hat{a}^{\dagger}_{\mathbf{k},\lambda}e^{i\omega t}\right]\nonumber \\
    & \qq{}\qq{}\qq{}\qq{}\qq{} \cdot (\mathbf{k} \times \text{e}_{\mathbf{k},\lambda})_j\,,\\ 
     \partial_i B_j  &= -\sum_{\mathbf{k},\lambda} \left(\frac{\hbar}{2\epsilon_0 \omega V}\right)^{1/2}k_i \left[ \hat{a}_{\mathbf{k},\lambda}e^{-i\omega t}+ \hat{a}^{\dagger}_{\mathbf{k},\lambda}e^{i\omega t}\right]\nonumber \\
     & \qq{}\qq{}\qq{}\qq{}\qq{} \cdot(\mathbf{k} \times \text{e}_{\mathbf{k},\lambda})_j \,.
\end{align}
Here, $V$ is the volume of momentum space, $\omega$ is the frequency related to the wavenumber $\abs{\mathbf{k}}$ and $\text{e}_{\mathbf{k},\lambda}$ are the orthogonal polarisation vectors corresponding to the two polarisations given by $\lambda = 1,2$; they satisfy the conditions:
 $  \text{e}_{\mathbf{k},\lambda_1} \cdot \text{e}_{\mathbf{k},\lambda_2} = \delta_{\lambda_1,\lambda_2}$ and $\text{e}_{\mathbf{k},\lambda} \cdot \mathbf{k} = 0 $~\cite{MILONNI_2023}.
\textcolor{black}{Moreover, the term \((\mathbf{k} \times \text{e}_{\mathbf{k},\lambda})_j\) denotes the \(j\)-th component of the cross product.}
The annihilation and the creation operators, $\hat{a}_{\mathbf{k},\lambda},~\hat{a}_{\mathbf{k},\lambda}^{\dagger}$ satisfy the standard commutation relationship of a boson. Furthermore, they can be related to the occupation number, which can be recast in terms of the Bose-Einstein distribution in a finite temperature case.
\begin{align}
    \langle \hat{a}_{\mathbf{k},\lambda}\hat{a}_{\mathbf{k},\lambda}^\dag\rangle&=n(\omega)+1\,,\label{quant operator1}\\
    \langle \hat{a}_{\mathbf{k},\lambda}^\dag \hat{a}_{\mathbf{k},\lambda}\rangle &= \langle \hat{a}_{\mathbf{k},\lambda}\hat{a}_{\mathbf{k},\lambda}^\dag\rangle -1 =n(\omega) \label{quant operator2}\,,
\end{align}
where 
\begin{align}
\textcolor{black}{n(\omega)=\frac{1}{(e^{\hbar\omega/k_B T}-1)}}\,,
\end{align}
and where $k_B$ is the Boltzmann constant and $T$ the ambient temperature.
In the static limit, which we are working in here, to obtain the momentum diffusion constant, we need to find the contribution of the magnetic dipole moment and its magnetic field to the impulse, $\Delta p_i$, see Eq.~\eqref{eq:deltapi}.
According to the statistical independence described in appendix \ref{section: appendix} and taking the continuous limit, i.e.
$
    \sum_{\mathbf{k}} \rightarrow \frac{V}{(2\pi)^3} \int \dd[3]{k}. 
$
We want to resolve the integral by integrating over the full momentum space.
Taking the average via $\langle \Delta p^2 \rangle = \frac{1}{3} \sum_{i= x,y,z} \langle \Delta p_i^2 \rangle \,$, and using Eqs.~\eqref{quant operator1},~\eqref{quant operator2}, we obtain $\langle \Delta p^2\rangle$ as shown in Eq.~\eqref{eq:ave_p} by following the commutation relationship between annihilation and creation operators.
\begin{widetext}
\begin{align}
    \begin{split}
        \Delta p_i &=-2i\alpha \sum_{\mathbf{k}_1,\lambda_1}\sum_{\mathbf{k}_2,\lambda_2}\left(\frac{\hbar}{2\epsilon_0\omega_2 V}\right)^{1/2}  \left(\frac{\hbar}{2\epsilon_0\omega_1 V}\right)^{1/2} k_{1,i}
        \frac{\sin(1/2(\omega_1-\omega_2)\Delta t)}{\omega_1-\omega_2}\\
        &\quad \cdot \Biggl[\hat{a}_{\mathbf{k}_2,\lambda_2}\hat{a}_{\mathbf{k}_1,\lambda_1}^\dagger e^{i(\omega_1-\omega_2)\Delta t/2} -\hat{a}^\dagger_{\mathbf{k}_2,\lambda_2}\hat{a}_{\mathbf{k}_1,\lambda_1}e^{-i(\omega_1-\omega_2)\Delta t/2}\Biggl]
        ((\mathbf{k}_2 \times \text{e}_{\mathbf{k}_2,\lambda_2})_j) ((\mathbf{k}_1 \times \text{e}_{\mathbf{k}_1,\lambda_1})_j)\,, \label{eq:deltapi}
    \end{split}\\
    \begin{split}
    \langle \Delta p ^2 \rangle
        &= \frac{1}{3} \frac{8\hbar^2}{4\epsilon_0^2(8\pi^3)^2c^{8}}\frac{4\pi}{3}4\pi \alpha^2 \int_0^{\infty}d\omega_1\omega_1^3  \int_0^{\infty}d\omega_2\omega_2 \frac{\sin^2(1/2(\omega_1-\omega_2)\Delta t)}{(\omega_1-\omega_2)^2} \\
       &\qq{}\cdot
       [(n(\omega_2)+1)n(\omega_1) +(n(\omega_1)+1)n(\omega_2) ]
       \left(\sum_{\lambda_1} \text{e}_{\textbf{k}_2 \lambda_{2}}^2 \right)\left(\sum_{\lambda_2} \text{e}_{\textbf{k}_1 \lambda_{1}}^2\right) \label{eq:ave_p}
       \end{split} \\
       \begin{split}
       &= \frac{2 \hbar^2 \alpha^2 }{9\pi^4 \epsilon_0^2 c^{12}} \int_0^{\infty}d\omega\omega^{8}
    [ n(\omega)^2 + n(\omega)]\frac{\pi\Delta t}{2} \, .
    \end{split} \label{eqn: k to omega}
\end{align}
\end{widetext}
To reach Eq.~\eqref{eq:ave_p} we used $ |\mathbf{k}|c=\omega $ to change the variable in the integration,
and we resolve one integral via the substitution $x=\Delta t(\omega_1-\omega_2)/2$: 
\begin{align}
    &\int_0^{\infty} \text{d}\omega_2\, \frac{\sin^2(1/2(\omega_1-\omega_2)\Delta t)}{(\omega_1-\omega_2)^2} \nonumber \\
    &\qq{} \approx \frac{1}{2} \int_{-\infty}^{\infty} \text{d}x\, \frac{\Delta t \sin^2 x}{x^2}
    = \frac{\pi \Delta t}{2}\,.
\end{align}
Moreover, we use the polarization summation:
\begin{eqnarray} \label{eq:orthogonal}   
    \left(\sum_{\lambda_2} \text{e}_{\textbf{k}_2 ,\lambda_{2}}^2 \right)\left(\sum_{\lambda_1} \text{e}_{\textbf{k}_1, \lambda_{1}}^2\right) &= 2\cdot 2 = 4\,,
\end{eqnarray}
where the summations only count orthogonal directions to $\mathbf{k}$. The relation in Eq.~\eqref{eq:orthogonal} is used to resolve the terms originating from the cross-product when deriving $\Delta p_{i}^2$, such as:
\begin{eqnarray}
    (\mathbf{k}_2 \times \text{e}_{\mathbf{k}_2,\lambda_2})^2  = ({\textbf{k}_2} \cdot {\textbf{k}_2}) (\text{e}_{\textbf{k}_2 \lambda_{2}} \cdot \text{e}_{\textbf{k}_2 \lambda_{2}}).
\end{eqnarray}
\textcolor{black}{Moreover, the intermediate steps between squaring and integration are provided in Appendix \ref{section: appendix}.} For the magnetic field fluctuations at finite ambient temperature, $T$, which are assumed to be in thermal equilibrium, we obtain~\footnote{At this point, it is worth comparing with a point dipole case. The analogous expression for a dipole has been derived earlier, ${\Delta p^2}/{\Delta t} =  \frac{8\hbar^2}{3 \pi c^5}  \int_0^\infty \text{d}\omega\, \omega^8 |\alpha (\omega)|^2 [n^2(\omega)+ n(\omega)]$, see~\cite{sinha_dipoles_2022}.}:
\begin{align}
    \frac{\langle \Delta p ^2 \rangle}{\Delta t} &= \frac{ \hbar^2 \alpha^2 }{9 \pi^3 \epsilon_0^2 c^{12}} \int_0^{\infty}d\omega\omega^{8}
    [n^2(\omega)+n(\omega)]. \label{eqn: diamagnetism result}
\end{align}
\textcolor{black}{Let us now consider an example of a magnetic sphere with radius, $a$, and magnetic susceptibility $\chi_v$} (which relates to the permeability $\mu$ via $\mu \equiv \mu_0 (1+ \chi_v)$~\cite{Zangwill_2012}); denoting the vacuum permeability with $\mu_0$, its magnetic moment induced by an external magnetic field $\mathbf{B}_0$ is given by: 
\begin{align}
    \mathbf{m} &=  \frac{a^3}{\mu_0} \frac{\mu-\mu_0}{\mu + 2\mu_0} \mathbf{B}_0\,.
\end{align}
From Eq.~\eqref{eq:mj} we thus have:
\begin{align}
    \alpha &= \frac{a^3}{\mu_0} \frac{\mu-\mu_0}{\mu + 2\mu_0} \,.
\end{align}
The momentum fluctuation of the nanoparticle due to magnetic field fluctuations, using $c^2 = \epsilon_0 \mu_0$, are therefore found to be:
\begin{align}
     \frac{\langle \Delta p^2 \rangle}{\Delta t} = 
      \frac{ a^6 \hbar^2 c}{9\pi^3 } \abs{ \frac{\mu-\mu_0}{\mu + 2\mu_0} }^2 \left( \frac{k_B T}{c \hbar}\right)^{9} \zeta (8) \Gamma (8)\,, \label{eqn: magnetic sphere result}
\end{align}
where $\zeta (8)=\pi^8/9450\sim 1$ is the Riemann zeta function, and 
$\Gamma(8)$ is the Gamma function.
A similar temperature dependence of $T^{9}$ for the momentum diffusion term was obtained for a dielectric sphere in the presence of the electric field fluctuations~\cite{sinha_dipoles_2022}, which will be discussed below.
Note that the result in Eq.~\eqref{eqn: magnetic sphere result} takes into account both the $n$ and $n^2$-term, while often the $n^2$-term is neglected due to its small contribution in quantum systems~\cite{sinha_dipoles_2022}.
The absolute sign is due to the square of the, possibly complex, $\alpha$~\footnote{\textcolor{black}{The polarizability depends on the magnetic permeability \( \mu \), which can be a complex number. However, in the case of nanodiamonds, the permeability \(\chi_v\) is real, making \( \mu \) effectively real, see Refs.~\cite{richards_time-resolved_2025, yelisseyev_magnetic_2009, poklonski_magnetic_2023}. Consequently, we treat the polarizability as a real quantity.}}.

\section{Decoherence rate}

It was shown in Ref.~\cite{sinha_dipoles_2022} that the decoherence rate can be related to the momentum diffusion constant; it produces the decoherence rate due to the electric field fluctuations in a thermal bath. 
Similarly, we now compute the decoherence rate induced by the magnetic field fluctuations for a diamagnetic sphere. 
We present the decoherence rate in terms of the volume susceptibility of the diamagnetic material:
\begin{align}
   \textcolor{black}{\gamma_{\rm B}} &= \Lambda (\Delta x)^2 = \frac{1}{2\hbar^2}\frac{\langle \Delta p^2 \rangle}{\Delta t} (\Delta x)^2 \nonumber \\
    &= \frac{ a^6 c }{18 \pi^3} \abs{\frac{\chi_v}{3+\chi_v} }^2 \left( \frac{k_B T}{c\hbar}\right)^{9} \zeta (8) \Gamma (8) (\Delta x)^2 \,,\label{eqn: lambda of dia}
\end{align}
where we have used the volume magnetic susceptibility $\chi_v$ rather than the magnetic permeability $\mu$.
Since the momentum diffusion rate computed here is a mean-square ensemble average over momentum fluctuations, it has no direct bearing on the entanglement between the nanoparticle and the ambient magnetic field fluctuations, as first pointed out in Ref.~\cite{sinha_dipoles_2022}. 
However, it was also mentioned that the analogous computation for the dielectric case in the presence of electric field fluctuations provides us with the decoherence result, which is consistent with the rigorous calculations of the decoherence rate obtained in~\cite{schlosshauer,Hornberger:2003,adler_normalization_2006,Fragolino:2023agd,Chang_2009} for a spatial superposition. 
Previously, in Ref.~\cite{Adler:2006gt}, the author also compared a diffusion rate with the results of collisional-based decoherence.

Now, let us compare this decoherence rate with the \textcolor{black}{one due to the electric field in a thermal bath}. 
As stated above, this has been previously studied rigorously in Ref.~\cite{sinha_dipoles_2022}, and was found to be given by
\begin{equation}\label{Lambda-electric}
\textcolor{black}{\gamma_{\rm E}}=\Lambda (\Delta x)^2 = \frac{512\pi^7 a^6 c}{135} \left(\frac{k_BT}{\hbar c}\right)^9\left|\frac{\epsilon-1}{\epsilon+2}\right|^2 (\Delta x)^2 \, ,
\end{equation}
where $\epsilon$ is the dielectric constant. The ratio of the two decoherence rates (magnetic over electric) is then \textit{solely} given by the material properties:
\begin{equation}\label{ratio}
\textcolor{black}{\frac{\gamma_{\rm B}}{\gamma_{\rm E}}}=\frac{135}{19216\times \pi^{10}}\abs{\frac{\chi_v}{3+\chi_v} }^2
\left|\frac{\epsilon+2}{\epsilon-1}\right|^2 \, .
\end{equation}
For typical solid material, $\epsilon> 1$, and for diamagnetic material, $\chi_v \ll 1$. 
For a pure superconductor $\chi_v=-1$; in this case, the above ratio is very suppressed: $\sim10^{-7}$ for $\epsilon\gg 1$.
To take an example of a diamagnetic material: for diamond $\chi_v\sim -2.2\times 10^{-5}$, while $\epsilon\sim 5.7$. 
Hence, this ratio is $\textcolor{black}{\gamma_{\rm B}/\gamma_{\rm E}}\sim 10^{-17}$, suggesting that the decoherence rate for any spatial superposition due to magnetic field fluctuations is utterly negligible in the case of nanodiamonds. 
Hence, the dominant decoherence will arise from the dielectric properties of the material and other collisional and electromagnetic sources, see \cite{Schut:2024lgp,Schut:2023tce,Fragolino:2023agd}.

As a special case for testing the quantum nature of gravity in a lab, the electric-field-induced decoherence in Eq.~\eqref{Lambda-electric} has already been investigated for the case of nanodiamonds of mass $m\sim 10^{-14}$~kg; see~\cite{RomeroIsart2011LargeQS, Rijavec:2020qxd}. 
\textcolor{black}{By considering Eq.~\eqref{Lambda-electric}, the required ambient temperature is $T\sim {\cal O}(5)$~K for decoherence rate $\gamma \leq 10^{-1}\,\si{\hertz}$ and for $\Delta x\sim 11 {\rm \mu m}$, where the thermal wavelength is on the order of millimeters at this temperature, which corresponds to the validity of the long-wavelength approximation in this regime.}
As mentioned above, the magnetic contribution to the decoherence rate in this example will be negligible. 

In summary, we have found that the magnetic field fluctuations in a thermal bath on a diamagnetic material are negligible.
However, we must also adhere to the decoherence bound by the electric field fluctuations occurring in the vacuum and in a thermal bath, which poses much stricter bounds in most cases.

\section*{Acknowledgements}
AM's research is funded by the Gordon and Betty Moore Foundation through Grant GBMF12328, DOI 10.37807/GBMF12328, and  Alfred P. Sloan Foundation under Grant No. G-2023-21130. We are thankful to K. Sinha and P. Milonni for their helpful discussions.

\nocite{}

\bibliography{reference}

\begin{thebibliography}{10}

\bibitem{simon_diamagnetic_2000}
M.~D. Simon and A.~K. Geim, ``Diamagnetic levitation: {Flying} frogs and floating magnets (invited),'' {\em Journal of Applied Physics}, vol.~87, pp.~6200--6204, May 2000.

\bibitem{Simon:2001}
M.~Simon, L.~Heflinger, T.~Ca, and A.~Geim, ``Diamagnetically stabilized magnet levitation,'' {\em American Journal of Physics}, vol.~69, 06 2001.

\bibitem{Nakashima:2020jnr}
R.~Nakashima, ``{Diamagnetic levitation of a milligram-scale silica using permanent magnets for the use in a macroscopic quantum measurement},'' {\em Phys. Lett. A}, vol.~384, p.~126592, 2020.

\bibitem{hunter2023diamagnetic}
G.~Hunter-Brown, {\em Diamagnetic Levitation of Bubbles and Droplets}.
\newblock PhD thesis, University of Nottingham, 2023.

\bibitem{hunter2023sonomaglev}
G.~Hunter-Brown, N.~Sampara, M.~M. Scase, and R.~J. Hill, ``Sonomaglev: Combining acoustic and diamagnetic levitation,'' {\em Applied Physics Letters}, vol.~122, no.~1, 2023.

\bibitem{Ashkarran_2024}
A.~A. Ashkarran and M.~Mahmoudi, ``Magnetic levitation of nanoscale materials: the critical role of effective density,'' {\em Journal of Physics D: Applied Physics}, vol.~57, p.~065001, nov 2023.

\bibitem{Hofer:2022chf}
J.~Hofer {\em et~al.}, ``{High-Q Magnetic Levitation and Control of Superconducting Microspheres at Millikelvin Temperatures},'' {\em Phys. Rev. Lett.}, vol.~131, no.~4, p.~043603, 2023.

\bibitem{chen2022diamagnetic}
X.~Chen, S.~K. Ammu, K.~Masania, P.~G. Steeneken, and F.~Alijani, ``Diamagnetic composites for high-q levitating resonators,'' {\em Advanced Science}, vol.~9, no.~32, p.~2203619, 2022.

\bibitem{tian2023feedbackcoolinginsulatinghighq}
S.~Tian, K.~Jadeja, D.~Kim, A.~Hodges, G.~C. Hermosa, C.~Cusicanqui, R.~Lecamwasam, J.~E. Downes, and J.~Twamley, ``Feedback cooling of an insulating high-q diamagnetically levitated plate,'' {\em Applied Physics Letters}, vol.~124, no.~12, p.~124002, 2024.

\bibitem{Qian:2013}
A.~R. Qian, D.~C. Yin, P.~F. Yang, Y.~Lv, Z.~C. Tian, and P.~Shang, ``Application of diamagnetic levitation technology in biological sciences research,'' {\em IEEE Transactions on Applied Superconductivity}, vol.~23, no.~1, pp.~3600305--3600305, 2013.

\bibitem{schilling2021physicsdiamagneticlevitation}
A.~Schilling, ``The physics of diamagnetic levitation,'' 2021.
\newblock arXiv:2101.02160 [physics.ed-ph].

\bibitem{Elahi:2024dbb}
S.~G. Elahi, M.~Schut, A.~Dana, A.~Grinin, S.~Bose, A.~Mazumdar, and A.~Geraci, ``Diamagnetic micro-chip traps for levitated nanoparticle entanglement experiments,'' 2024.
\newblock ArXiv:2411.02325.

\bibitem{van_de_kamp_quantum_2020}
T.~W. van~de Kamp, R.~J. Marshman, S.~Bose, and A.~Mazumdar, ``Quantum gravity witness via entanglement of masses: Casimir screening,'' {\em Phys. Rev. A}, vol.~102, no.~6, p.~062807, 2020.

\bibitem{Hsu:2016}
J.~Hsu, P.~Ji, C.~W. Lewandowski, and B.~D’Urso, ``Cooling the motion of diamond nanocrystals in a magneto-gravitational trap in high vacuum,'' {\em Scientific reports}, vol.~6, no.~1, p.~30125, 2016.

\bibitem{Slezak_2018}
B.~R. Slezak, C.~W. Lewandowski, J.-F. Hsu, and B.~D'Urso, ``Cooling the motion of a silica microsphere in a magneto-gravitational trap in ultra-high vacuum,'' {\em New Journal of Physics}, vol.~20, p.~063028, jun 2018.

\bibitem{Doherty_2013}
M.~W. Doherty, N.~B. Manson, P.~Delaney, F.~Jelezko, J.~Wrachtrup, and L.~C. Hollenberg, ``The nitrogen-vacancy colour centre in diamond,'' {\em Physics Reports}, vol.~528, p.~1–45, July 2013.

\bibitem{Greve:2021wil}
G.~P. Greve, C.~Luo, B.~Wu, and J.~K. Thompson, ``{Entanglement-enhanced matter-wave interferometry in a high-finesse cavity},'' {\em Nature}, vol.~610, no.~7932, pp.~472--477, 2022.

\bibitem{Wan_2016}
C.~Wan, M.~Scala, G.~Morley, A.~A. Rahman, H.~Ulbricht, J.~Bateman, P.~Barker, S.~Bose, and M.~Kim, ``Free nano-object ramsey interferometry for large quantum superpositions,'' {\em Physical Review Letters}, vol.~117, Sept. 2016.

\bibitem{Nair:2023ovu}
S.~R. Nair, S.~Tian, G.~K. Brennen, S.~Bose, and J.~Twamley, ``Massive quantum superpositions using magnetomechanics,'' {\em Physical Review Applied}, 2025.
\newblock Accepted 20 June 2025.

\bibitem{Scala_2013}
M.~Scala, M.~S. Kim, G.~W. Morley, P.~F. Barker, and S.~Bose, ``Matter-wave interferometry of a levitated thermal nano-oscillator induced and probed by a spin,'' {\em Physical Review Letters}, vol.~111, Oct. 2013.

\bibitem{PhysRevLett.125.023602}
J.~S. Pedernales, G.~W. Morley, and M.~B. Plenio, ``Motional dynamical decoupling for interferometry with macroscopic particles,'' {\em Phys. Rev. Lett.}, vol.~125, p.~023602, Jul 2020.

\bibitem{Folman:2013}
S.~Machluf, Y.~Japha, and R.~Folman, ``Coherent stern--gerlach momentum splitting on an atom chip,'' {\em Nature communications}, vol.~4, p.~2424, 09 2013.

\bibitem{PinoEtAl2016}
H.~Pino, J.~Prat-Camps, K.~Sinha, B.~P. Venkatesh, and O.~Romero-Isart, ``On-chip quantum interference of a superconducting microsphere,'' {\em Quantum Science and Technology}, vol.~3, no.~2, p.~025001, 2018.

\bibitem{marshman_constructing_2022}
R.~J. Marshman, A.~Mazumdar, R.~Folman, and S.~Bose, ``Constructing {Nano}-{Object} {Quantum} {Superpositions} with a {Stern}-{Gerlach} {Interferometer},'' {\em Physical Review Research}, vol.~4, p.~023087, May 2022.

\bibitem{PhysRevLett.123.083601}
O.~Amit, Y.~Margalit, O.~Dobkowski, Z.~Zhou, Y.~Japha, M.~Zimmermann, M.~A. Efremov, F.~A. Narducci, E.~M. Rasel, W.~P. Schleich, and R.~Folman, ``${T}^{3}$ stern-gerlach matter-wave interferometer,'' {\em Phys. Rev. Lett.}, vol.~123, p.~083601, Aug 2019.

\bibitem{wu:2022rdv}
M.~Wu, M.~Toro\v{s}, S.~Bose, and A.~Mazumdar, ``{Quantum gravitational sensor for space debris},'' {\em Phys. Rev. D}, vol.~107, no.~10, p.~104053, 2023.

\bibitem{Debuisschert_2021}
T.~Debuisschert, ``Quantum sensing with nitrogen-vacancy colour centers in diamond,'' {\em Photoniques}, p.~50–54, Mar 2021.

\bibitem{Bose:2022czr}
S.~Bose, A.~Mazumdar, M.~Schut, and M.~Toro\v{s}, ``{Entanglement Witness for the Weak Equivalence Principle},'' {\em Entropy}, vol.~25, no.~3, p.~448, 2023.

\bibitem{Barker:2022mdz}
P.~F. Barker, S.~Bose, R.~J. Marshman, and A.~Mazumdar, ``{Entanglement based tomography to probe new macroscopic forces},'' {\em Phys. Rev. D}, vol.~106, no.~4, p.~L041901, 2022.

\bibitem{Elahi:2023ozf}
S.~G. Elahi and A.~Mazumdar, ``{Probing massless and massive gravitons via entanglement in a warped extra dimension},'' {\em Phys. Rev. D}, vol.~108, no.~3, p.~035018, 2023.

\bibitem{Vinckers:2023grv}
U.~K. Beckering~Vinckers, {\'A}.~De~La Cruz-Dombriz, and A.~Mazumdar, ``Quantum entanglement of masses with nonlocal gravitational interaction,'' {\em Physical Review D}, vol.~107, no.~12, p.~124036, 2023.

\bibitem{Chakraborty:2023kel}
S.~Chakraborty, A.~Mazumdar, and R.~Pradhan, ``{Distinguishing Jordan and Einstein frames in gravity through entanglement},'' {\em Phys. Rev. D}, vol.~108, no.~12, p.~L121505, 2023.

\bibitem{Bose:2017nin}
S.~Bose, A.~Mazumdar, G.~W. Morley, H.~Ulbricht, M.~Toro\v{s}, M.~Paternostro, A.~Geraci, P.~Barker, M.~S. Kim, and G.~Milburn, ``{Spin Entanglement Witness for Quantum Gravity},'' {\em Phys. Rev. Lett.}, vol.~119, no.~24, p.~240401, 2017.

\bibitem{ICTS}
\url{https://www.youtube.com/watch?v=0Fv-0k13s_k}, 2016.
\newblock Accessed 11/08/25.

\bibitem{Marletto:2017kzi}
C.~Marletto and V.~Vedral, ``{Gravitationally-induced entanglement between two massive particles is sufficient evidence of quantum effects in gravity},'' {\em Phys. Rev. Lett.}, vol.~119, no.~24, p.~240402, 2017.

\bibitem{Deli__2020}
U.~Deli{\'{c}}, M.~Reisenbauer, K.~Dare, D.~Grass, V.~Vuleti{\'{c}}, N.~Kiesel, and M.~Aspelmeyer, ``Cooling of a levitated nanoparticle to the motional quantum ground state,'' {\em Science}, vol.~367, pp.~892--895, feb 2020.

\bibitem{Piotrowski_2023}
J.~Piotrowski, D.~Windey, J.~Vijayan, C.~Gonzalez-Ballestero, A.~de~los R{\'{\i}}os~Sommer, N.~Meyer, R.~Quidant, O.~Romero-Isart, R.~Reimann, and L.~Novotny, ``Simultaneous ground-state cooling of two mechanical modes of a levitated nanoparticle,'' {\em Nature Physics}, vol.~19, pp.~1009--1013, mar 2023.

\bibitem{rytov_theory_1959}
S.~M. Rytov, {\em Theory of {Electric} {Fluctuations} and {Thermal} {Radiation}}.
\newblock Air Force Cambridge Research Center, 1959.

\bibitem{zheng_review_2014}
Y.~Zheng, ``Review of fluctuational electrodynamics and its applications to radiative momentum, energy and entropy transport,'' Oct. 2014.
\newblock arXiv:1410.5741.

\bibitem{brevik_fluctuational_2022}
I.~Brevik, B.~Shapiro, and M.~Silveirinha, ``Fluctuational electrodynamics in and out of equilibrium,'' {\em International Journal of Modern Physics A}, vol.~37, p.~2241012, July 2022.

\bibitem{sinha_dipoles_2022}
K.~Sinha and P.~W. Milonni, ``Dipoles in blackbody radiation: momentum fluctuations, decoherence, and drag force,'' {\em Journal of Physics B: Atomic, Molecular and Optical Physics}, vol.~55, p.~204002, Oct. 2022.

\bibitem{Fokker:1914}
A.~D. {Fokker}, ``{Die mittlere Energie rotierender elektrischer Dipole im Strahlungsfeld},'' {\em Annalen der Physik}, vol.~348, pp.~810--820, Jan. 1914.

\bibitem{Joos_Zeh_1985}
E.~Joos and H.~D. Zeh, ``The emergence of classical properties through interaction with the environment,'' {\em European Physical Journal B}, vol.~59, p.~223–243, June 1985.

\bibitem{Berg-Sorenson_1992}
K.~Berg-Sorenson, Y.~Castin, E.~Bonderup, and K.~Molmer, ``Momentum diffusion of atoms moving in laser fields,'' {\em Journal of Physics B: Atomic, Molecular and Optical Physics}, vol.~25, p.~4195, oct 1992.

\bibitem{Ghirardi}
G.~C. Ghirardi, A.~Rimini, and T.~Weber, ``Unified dynamics for microscopic and macroscopic systems,'' {\em Phys. Rev. D}, vol.~34, pp.~470--491, Jul 1986.

\bibitem{Balykin_1986}
V.~I. Balykin and A.~I. Sidorov, ``Optical pressure force in traveling and standing light waves during excitation of a multilevel atom by two-frequency laser radiation,'' {\em Soviet Journal of Quantum Electronics}, vol.~16, p.~1487, nov 1986.

\bibitem{Dalibard_1985}
J.~Dalibard and C.~Cohen-Tannoudji, ``Atomic motion in laser light: connection between semiclassical and quantum descriptions,'' {\em Journal of Physics B: Atomic and Molecular Physics}, vol.~18, p.~1661, apr 1985.

\bibitem{Agarwal:1993}
G.~S. Agarwal and K.~M{\o}lmer, ``Correlation-function approach to the momentum diffusion of atoms moving in standing waves,'' {\em Phys. Rev. A}, vol.~47, pp.~5158--5164, Jun 1993.

\bibitem{MILONNI_2023}
P.~Milonni, {\em Introduction to Quantum Optics and quantum fluctuations}.
\newblock OXFORD UNIV PRESS US, 2023.

\bibitem{oxenius_kinetic_2012}
J.~Oxenius, {\em Kinetic {Theory} of {Particles} and {Photons}: {Theoretical} {Foundations} of {Non}-{LTE} {Plasma} {Spectroscopy}}.
\newblock Springer {Series} in {Electronics} and {Photonics}, Springer Berlin Heidelberg, 2012.

\bibitem{schlosshauer}
M.~Schlosshauer, {\em Decoherence: the Quantum-to-Classical Transition}.
\newblock Springer Berlin, Heidelberg: Springer, 2007.
\newblock The Frontiers Collection.

\bibitem{cheng_long-range_1999}
C.~C. Cheng and M.~G. Raymer, ``Long-{Range} {Saturation} of {Spatial} {Decoherence} in {Wave}-{Field} {Transport} in {Random} {Multiple}-{Scattering} {Media},'' {\em Physical Review Letters}, vol.~82, pp.~4807--4810, June 1999.

\bibitem{chklovskii_relaxation_1992}
D.~B. Chklovskii and P.~A. Lee, ``Relaxation of nuclear spin in atomic hydrogen due to long-range orbital currents in metal walls,'' {\em Physical Review B}, vol.~45, pp.~5240--5243, Mar. 1992.

\bibitem{nagaosa_experimental_1991}
N.~Nagaosa and P.~Lee, ``Experimental consequences of the uniform resonating-valence-bond state,'' {\em Physical Review B}, vol.~43, pp.~1233--1236, Jan. 1991.

\bibitem{BreuerPetruccione2002}
H.~P. Breuer and F.~Petruccione, {\em The Theory of Open Quantum Systems}.
\newblock Oxford, UK: Oxford University Press, first edition~ed., 2002.

\bibitem{Hornberger:2003}
K.~Hornberger and J.~E. Sipe, ``Collisional decoherence reexamined,'' {\em Phys. Rev. A}, vol.~68, p.~012105, Jul 2003.

\bibitem{Schut:2024lgp}
M.~Schut, P.~Andriolo, M.~Toroš, S.~Bose, and A.~Mazumdar, ``Expression for the decoherence rate due to air-molecule scattering in spatial qubits,'' {\em Physical Review A}, vol.~111, no.~4, p.~042211, 2025.

\bibitem{RomeroIsart2011LargeQS}
O.~Romero-Isart, A.~C. Pflanzer, F.~Blaser, R.~Kaltenbaek, N.~Kiesel, M.~Aspelmeyer, and J.~I. Cirac, ``Large quantum superpositions and interference of massive nanometer-sized objects,'' {\em Physical review letters}, vol.~107, no.~2, p.~020405, 2011.

\bibitem{Zangwill_2012}
A.~Zangwill, {\em Modern Electrodynamics}.
\newblock Cambridge University Press, 2012.

\bibitem{richards_time-resolved_2025}
B.~A. Richards, N.~Ristoff, J.~Smits, A.~J. Perez, I.~Fescenko, M.~D. Aiello, F.~Hubert, Y.~Silani, N.~Mosavian, M.~S. Ziabari, A.~Berzins, J.~T. Damron, P.~Kehayias, D.~Egbebunmi, J.~E. Shield, D.~L. Huber, A.~M. Mounce, M.~P. Lilly, T.~Karaulanov, A.~Jarmola, A.~Laraoui, and V.~M. Acosta, ``Time-resolved diamond magnetic microscopy of superparamagnetic iron-oxide nanoparticles,'' {\em ACS Nano}, vol.~19, pp.~10048--10058, Mar. 2025.

\bibitem{yelisseyev_magnetic_2009}
A.~P. Yelisseyev, V.~P. Afanasiev, and V.~N. Ikorsky, ``Magnetic susceptibility of natural diamonds,'' {\em Doklady Earth Sciences}, vol.~425, pp.~330--333, Mar. 2009.

\bibitem{poklonski_magnetic_2023}
N.~A. Poklonski, A.~A. Khomich, I.~A. Svito, S.~A. Vyrko, O.~N. Poklonskaya, A.~I. Kovalev, M.~V. Kozlova, R.~A. Khmelnitskii, and A.~V. Khomich, ``Magnetic and {Optical} {Properties} of {Natural} {Diamonds} with {Subcritical} {Radiation} {Damage} {Induced} by {Fast} {Neutrons},'' {\em Applied Sciences}, vol.~13, p.~6221, Jan. 2023.
\newblock Number: 10 Publisher: Multidisciplinary Digital Publishing Institute.

\bibitem{adler_normalization_2006}
S.~L. Adler, ``Normalization of collisional decoherence: squaring the delta function, and an independent cross-check,'' {\em Journal of Physics A: Mathematical and General}, vol.~39, p.~14067, Oct. 2006.

\bibitem{Fragolino:2023agd}
P.~Fragolino, M.~Schut, M.~Toro\v{s}, S.~Bose, and A.~Mazumdar, ``{Decoherence of a matter-wave interferometer due to dipole-dipole interactions},'' {\em Phys. Rev. A}, vol.~109, no.~3, p.~033301, 2024.

\bibitem{Chang_2009}
D.~E. Chang, C.~A. Regal, S.~B. Papp, D.~J. Wilson, J.~Ye, O.~Painter, H.~J. Kimble, and P.~Zoller, ``Cavity opto-mechanics using an optically levitated nanosphere,'' {\em Proceedings of the National Academy of Sciences}, vol.~107, pp.~1005--1010, dec 2009.

\bibitem{Adler:2006gt}
S.~L. Adler, ``{Normalization of Collisional Decoherence: Squaring the Delta Function, and an Independent Cross-Check},'' {\em J. Phys. A}, vol.~39, pp.~14067--14074, 2006.

\bibitem{Schut:2023tce}
M.~Schut, H.~Bosma, M.~Wu, M.~Toro\v{s}, S.~Bose, and A.~Mazumdar, ``{Dephasing due to electromagnetic interactions in spatial qubits},'' {\em Phys. Rev. A}, vol.~110, no.~2, p.~022412, 2024.

\bibitem{Rijavec:2020qxd}
S.~Rijavec, M.~Carlesso, A.~Bassi, V.~Vedral, and C.~Marletto, ``{Decoherence effects in non-classicality tests of gravity},'' {\em New J. Phys.}, vol.~23, no.~4, p.~043040, 2021.

\end{thebibliography}

\appendix
\section{\textcolor{black}{Detailed derivation of momentum diffusion}}\label{section: appendix}

{\color{black}
The intermediate steps from Eq.~\ref{eq:deltapi} to \ref{eqn: k to omega} can be simplified by the argument of statistical independence between \( B_j \) and \( \partial_i B_j \), following the approach presented in the Appendix of Ref.~\cite{sinha_dipoles_2022}. This has been proven for two Gaussian random processes, \( X \) and \( Y \), satisfying \( \langle X \rangle = \langle Y \rangle = \langle XY \rangle = 0 \), the joint probability distribution can be factorized as \( p_{XY}(X,Y) = p_X(X) p_Y(Y) \). In our case, both the magnetic field \( B_j \) and its spatial derivative \( \partial_i B_j \) are Gaussian processes that fulfill the conditions \( \langle B_j \rangle = \langle \partial_i B_j \rangle = \langle B_j \, \partial_i B_j \rangle = 0 \)~\cite{MILONNI_2023}. Therefore, they are statistically independent, allowing us to square the contributions from \(B_j\) and \(\partial_i B_j\) separately below.
Consequently, we obtain:
\begin{align}
    \begin{split}
        \left\langle \Delta p_i^2 \right\rangle &= 4 \alpha^2 \sum_{\bold{k}_1\lambda_1}\sum_{\bold{k}_2\lambda_2}\left(\frac{\hbar}{2\omega_2 \epsilon_0 V}\right)  \left(\frac{\hbar}{2\omega_1 \epsilon_0 V}\right) k^2_{1i}\\
        &\quad \cdot \frac{\sin^2(1/2(\omega_1-\omega_2)\Delta t)}{(\omega_1-\omega_2)^2}\\
        &\quad \cdot  [-\langle \hat{a}_{\mathbf{k}_2,\lambda_2}\hat{a}_{\mathbf{k}_2,\lambda_2} \hat{a}_{\mathbf{k}_1,\lambda_1}^\dagger \hat{a}_{\mathbf{k}_1,\lambda_1}^\dagger \rangle  e^{i(\omega_1-\omega_2)\Delta t}\\
        &\quad + \langle \hat{a}_{\bold{k}_2\lambda_2}\hat{a}_{\bold{k}_2\lambda_2}^\dagger\hat{a}_{\bold{k}_1\lambda_1}^\dagger\hat{a}_{\bold{k}_1\lambda_1} \rangle \\
        &\quad + \langle \hat{a}^\dagger_{\bold{k}_2\lambda_2}\hat{a}_{\bold{k}_2\lambda_2}\hat{a}_{\bold{k}_1\lambda_1}\hat{a}_{\bold{k}_1\lambda_1}^\dagger \rangle\\
        &\quad - \langle \hat{a}^\dagger_{\mathbf{k}_2,\lambda_2}\hat{a}_{\mathbf{k}_1,\lambda_1} \hat{a}^\dagger_{\mathbf{k}_2,\lambda_2}\hat{a}_{\mathbf{k}_1,\lambda_1} \rangle e^{-i(\omega_1-\omega_2)\Delta t} ]\\
        &\quad \cdot [({\textbf{k}_2} \times \text{e}_{\textbf{k}_2 \lambda_{k_2}})_j({\textbf{k}_2} \times \text{e}_{\textbf{k}_2 \lambda_{k_2}})_j]\\
        &\quad \cdot [({\textbf{k}_1} \times \text{e}_{\textbf{k}_1 \lambda_{k_1}})_k({\textbf{k}_1} \times \text{e}_{\textbf{k}_1 \lambda_{k_1}})_k]
    \end{split} \label{eqn: appendix, a1} \\
    \begin{split}
        &= 4 \alpha^2 \sum_{\bold{k}_1\lambda_1}\sum_{\bold{k}_2\lambda_2}\left(\frac{\hbar}{2\epsilon_0 \omega_2 V}\right)  \left(\frac{\hbar}{2\epsilon_0 \omega_1 V}\right) k^2_{1i}\\
        &\quad \cdot \frac{\sin^2(1/2(\omega_1-\omega_2)\Delta t)}{(\omega_1-\omega_2)^2} [\langle \hat{a}_{\bold{k}_2\lambda_2}\hat{a}_{\bold{k}_2\lambda_2}^\dag\rangle \langle \hat{a}_{\bold{k}_1\lambda_1}^\dag\hat{a}_{\bold{k}_1\lambda_1}\rangle\\
        &\quad +\langle \hat{a}^\dag_{\bold{k}_2\lambda_2}\hat{a}_{\bold{k}_2\lambda_2}\rangle \langle \hat{a}_{\bold{k}_1\lambda_1}\hat{a}_{\bold{k}_1\lambda_1}^\dag\rangle ]\\
        &\quad \cdot [({\textbf{k}_2} \times \text{e}_{\textbf{k}_2 \lambda_{k_2}})_j({\textbf{k}_2} \times \text{e}_{\textbf{k}_2 \lambda_{k_2}})_j]\\
        &\quad \cdot [({\textbf{k}_1} \times \text{e}_{\textbf{k}_1 \lambda_{k_1}})_k({\textbf{k}_1} \times \text{e}_{\textbf{k}_1 \lambda_{k_1}})_k]
    \end{split} \label{eqn: appendix, a2}
\end{align}
The term \((\mathbf{k} \times \text{e}_{\mathbf{k},\lambda})_j\) denotes the \(j\)-th component of the cross product.
Moreover, using the commutation relations for the annihilation and creation operators
\begin{equation}
    [\hat{a}_{\bold{k}_i\lambda_i}, \hat{a}^\dag_{\bold{k}_j\lambda_j}] = \delta_{ij}
\end{equation}
the cross terms involving exponential functions vanish upon taking the expectation value, i.e., \( \langle a a \rangle = \langle a^\dag a^\dag \rangle = 0 \), and \( \langle a^\dag a \rangle \propto n(\omega) \). Then, taking the average $\langle \Delta p^2 \rangle = \frac{1}{3} \sum_{i= x,y,z} \langle \Delta p_i^2 \rangle \,$ we get
\begin{align}
    \begin{split}
        \left\langle \Delta p^2 \right\rangle &= \frac{4}{3} \alpha^2 \sum_{\bold{k}_1\lambda_1}\sum_{\bold{k}_2\lambda_2}\left(\frac{\hbar}{2\epsilon_0 \omega_2 V}\right)  \left(\frac{\hbar}{2\epsilon_0 \omega_1 V}\right) k^2_{1}\\
        &\quad \cdot \frac{\sin^2(1/2(\omega_1-\omega_2)\Delta t)}{(\omega_1-\omega_2)^2} [\langle \hat{a}_{\bold{k}_2\lambda_2}\hat{a}_{\bold{k}_2\lambda_2}^\dag\rangle \langle \hat{a}_{\bold{k}_1\lambda_1}^\dag\hat{a}_{\bold{k}_1\lambda_1}\rangle\\
        &\quad +\langle \hat{a}^\dag_{\bold{k}_2\lambda_2}\hat{a}_{\bold{k}_2\lambda_2}\rangle \langle \hat{a}_{\bold{k}_1\lambda_1}\hat{a}_{\bold{k}_1\lambda_1}^\dag\rangle ]\\
        &\quad \cdot ({\textbf{k}_2} \times \text{e}_{\textbf{k}_2 \lambda_{k_2}})^2 ({\textbf{k}_1} \times \text{e}_{\textbf{k}_1 \lambda_{k_1}})^2.
    \end{split}
\end{align}
Lastly, by taking the continuous limit, we arrive Eq.~\ref{eqn: k to omega}.
}
\vfill

\end{document}